\documentclass[aps,superscriptaddress,prc,twocolumn,nofootinbib]{revtex4}

\usepackage{amsmath,bm}
\usepackage{graphicx}
\usepackage{epstopdf}
\usepackage{subfigure}
\usepackage{epsfig}
\usepackage{amsmath,amssymb,amsfonts}
\usepackage{color}
\usepackage[utf8]{inputenc}
\usepackage{verbatim}
\usepackage{array}
\graphicspath{{figures/}} 

\begin{document}

\title{Diffusion of charm quarks in jets in high-energy heavy-ion collisions}

\author{Sa Wang}
\affiliation{Key Laboratory of Quark \& Lepton Physics (MOE) and Institute of Particle Physics,
 Central China Normal University, Wuhan 430079, China}
 
\author{Wei Dai}
\email{weidai@cug.edu.cn}
\affiliation{School of Mathematics and Physics, China University
of Geosciences, Wuhan 430074, China}

\author{Ben-Wei Zhang}
\email{bwzhang@mail.ccnu.edu.cn}
\affiliation{Key Laboratory of Quark \& Lepton Physics (MOE) and Institute of Particle Physics, Central China Normal University, Wuhan 430079, China}

\author{Enke Wang}
\affiliation{Key Laboratory of Quark \& Lepton Physics (MOE) and Institute of Particle Physics,
 Central China Normal University, Wuhan 430079, China}
\date{\today}


\begin{abstract}
The radial distribution of $D^0$ mesons in jets probes the diffusion of charm quark relative to the jet axis and provides a new perspective to study the interaction mechanisms between heavy quarks and the medium in the nucleus-nucleus collisions. The in-medium parton propagations are described by a Monte Carlo transport model which uses the next-to-leading order (NLO) plus parton shower (PS) event generator SHERPA as input and includes elastic (collisional) and inelastic (radiative) interaction for heavy quarks as well as light partons.  At low $D^0$ meson $p_T$, the radial distribution significantly shifts to larger radius indicating a strong diffusion effect which is consistent with the recent experimental data. We demonstrate that the angular deviation of charm quarks declines with $p_T$ and is very sensitive to the collisional more than radiative interaction at $p_T<5$~GeV. As predictions, we present the $D^0$ meson radial distribution in jets in p+p and $0-10\%$ Au+Au collisions at $\sqrt{s_{NN}}=200$~GeV at the RHIC, and also estimate the nuclear modification factor of charm jet in central Au+Au collisions at 200~GeV at the RHIC and central Pb+Pb collisions at $5.02$~TeV at the LHC.

\end{abstract}


\maketitle

\section{ Introduction }
\label{sec:intro}
The quark-gluon plasma (QGP) formed in the high-energy nucleus-nucleus collisions, both at the Relativistic Heavy Ion Collider (RHIC) and the Large Hadron Collider (LHC), provides an arena to study the Quantum Chromodynamics (QCD) under such extreme hot and dense deconfined state of nuclear matter. The strong interaction between the high-$p_T$ parton produced in the hard scattering with the medium, referred as ``jet quenching" effect~\cite{Gyulassy:2003mc, Qin:2015srf, Vitev:2008rz, Vitev:2009rd, CasalderreySolana:2010eh,Young:2011qx,He:2011pd,ColemanSmith:2012vr,Neufeld:2010fj,Zapp:2012ak,Ma:2013pha, Senzel:2013dta, Casalderrey-Solana:2014bpa,Milhano:2015mng,Chang:2016gjp,Majumder:2014gda, Chen:2016cof, Chien:2016led, Apolinario:2017qay,Connors:2017ptx,Zhang:2018urd,Zhang:2018ydt}, has been extensively investigated owing to its direct connection to the properties of the QGP. Especially, owing to the large mass ($M_Q \gg T$), the heavy flavor production is set by the initial hard scattering before the QGP formed in A+A collisions while the subsequent thermal generation in the hot and dense medium could be negligible~\cite{Andronic:2015wma,Dong:2019unq}. Hence the heavy quarks are witnesses of the entire bulk medium evolution and strongly interact with the constituents of the QCD matter, therefore being ideal probes to the properties of QGP.

The experimental measurements including the nuclear modification factor $R_{AA}$~\cite{Adamczyk:2014uip,Adam:2015sza,Sirunyan:2017xss} and the azimuthal anisotropy $v_2$~\cite{Abelev:2014ipa,Adamczyk:2017xur,Acharya:2017qps,Sirunyan:2017plt} for charmed hadrons in both Au-Au at the RHIC and Pb+Pb collisions at the LHC show a strong interaction between the charm quarks and the QGP formed in A+A collisions. A lot of theoretical calculations~\cite{vanHees:2007me,CaronHuot:2008uh,vanHees:2005wb,Djordjevic:2015hra,He:2014cla,Chien:2015vja,Kang:2016ofv,Cao:2011et,Moore:2004tg,Cao:2013ita,Cao:2015hia,Cao:2017hhk,Alberico:2013bza,Xu:2015bbz,Cao:2016gvr,Das:2016cwd,Ke:2018tsh} have been made to confront with these measurements, which greatly improve our understanding of the in-medium heavy quarks evolution. Heavy quarks are believed to undergo Brownian motion at low $p_T$ due to their large mass, and behave like light quarks at very high $p_T$~\cite{Cao:2013ita,Dong:2019unq}. Furthermore, it's generally assumed that the collisional energy loss should dominate up to $p_T\sim 5M_Q$, and then the radiative energy loss would become dominant when it goes to higher $p_T$ region where the mass effects disappear.

The heavy-flavour jet measurements such as $b$-jet~\cite{Chatrchyan:2013exa} and $b\bar{b}$ dijet~\cite{Sirunyan:2018jju} in Pb+Pb collisions also shed new light on the mass dependence of jet quenching~\cite{ Li:2017wwc,Li:2018xuv,Kang:2018wrs,Dai:2018mhw,Wang:2018gxz}. The recent reported $D^0$ meson radial profile in jets ($D^0$ meson inside the jets, denoting as $D^0-$jet below) both in p+p and $0-100\%$ Pb+Pb collisions at $\sqrt{s_{NN}}=5.02$~TeV measured by CMS collaboration~\cite{CMS:2018ovh} was regarded as a complementary observable to the previous measurements of heavy-flavour jets~\cite{Chatrchyan:2013exa,Sirunyan:2018jju}. Diffusion of low $p_T$ $D^0$ meson relative to a high $p_T$ jet provides a new perspective to study the dynamical details of the heavy quarks in-medium interaction and may act as a candidate to disentangle the collisional and radiative mechanism more than the charm quark energy loss, then help us to understand how the non-perturbative interactions (collisional process) of heavy quarks at low $p_T$ give way to the perturbative interaction (radiative process) at high $p_T$. Besides, the recent measured production of c-jet, which is defined as a jet containing charm quark inside the jet cone, both in p+p and p+A collisions~\cite{Aad:2011td,Sirunyan:2016fcs,Acharya:2019zup} can be exploited as a baseline to address the final state in-medium modification of charm jet production in A+A collisions.

With a Monte Carlo (MC) transport simulation combining elastic (collisional) and inelastic (radiative) interaction of the energetic heavy (light) quarks with the hot and dense medium, employing a next-to-leading order (NLO) plus parton shower (PS) generated initial events as input, we study the in-medium modification of the radial distribution of $D^0$ meson in jets both at the RHIC and the LHC. We find the net effects of collisional and radiative energy loss in the charm quarks radial diffusion in jets is quite different, providing a good chance to constrain the essential difference of the collisional and radiative energy loss. Furthermore, we will present our predictions of the nuclear modification factor of charm jet for the future experimental measurements.

The paper is organized as follows. In Sec.~II, we will show our p+p setups used in the MC simulation and confront our p+p baseline with the experimental results. The framework of in-medium parton energy loss both for heavy quark and light partons would be introduced in Sec.~III. In Sec.~IV, we will present our calculations of the radial distribution of $D^0$ meson in jets in p+p and Pb+Pb collisions and compare them with the recent preliminary experimental data. Also the predictions of the nuclear modification factor of the charm quark tagged jet both at the RHIC and the LHC will be given. A summary will be given in Sec.~V.

\section{Charm jet production and $D^0$ meson radial profile in jets in p+p collisions}
\label{sec:ppbaseline}
\par In this work, we use the event generator \textsl{SHERPA}, which computes the NLO QCD matrix elements matched with parton shower, to generate the inital parton-level c-jet events in p+p collisions as input of for Monte Carlo simulation. The NLO matrix elements are matched with the parton shower using the MC@NLO method~\cite{Frixione:2002ik}. The NNPDF 3.0 NNLO parton distribution function (PDF) with 4-flavor sets~\cite{Ball:2014uwa} have been chosen in the computation. The jet reconstruction and event selection are implemented with anti-$k_T$ algorithm~\cite{Cacciari:2008gp} within \textsl{FASTJET}~\cite{Cacciari:2011ma}. To compare with the experimental data, the fragmentation process $c(\bar{c})\rightarrow D^0$ of charm quark is performed based on the Peterson fragmentation functions (FFs)~\cite{Peterson:1982ak} $D(z)=N/z/(1-1/z-\epsilon_c/(1-z))^2$, where $N=0.183$, $\epsilon_c=0.01$ (which has been chosen in Refs.~\cite{Das:2016llg} to reprodue the D meson spectra in p+p collisions at LHC energy using FONLL ~\cite{Cacciari:2005rk,Cacciari:2012ny}) and $z$ refers to the longitudinal momentum fraction carried by the $D^0$ meson. And the branch ratio (BR) ($c\rightarrow D^0$)=0.168~\cite{Cacciari:2003zu} is used. We note that the fragmentation is the dominant process to the charm quark hadronization at $p_T > 5$~GeV, hence we neglect the colesence effect~\cite{Cao:2013ita,Song:2015sfa} in this study. We have not include the hadronization effect of light partons in this work. Because of the small modification on the c-jet production observed in p+A collisions~\cite{Sirunyan:2016fcs}, we ignore the Cold Nuclear Matter (CNM) effects in our current study.

\begin{figure}[!t]
\begin{center}
\vspace*{0.2in}
\hspace*{-.1in}
\subfigure[]{
  \epsfig{file=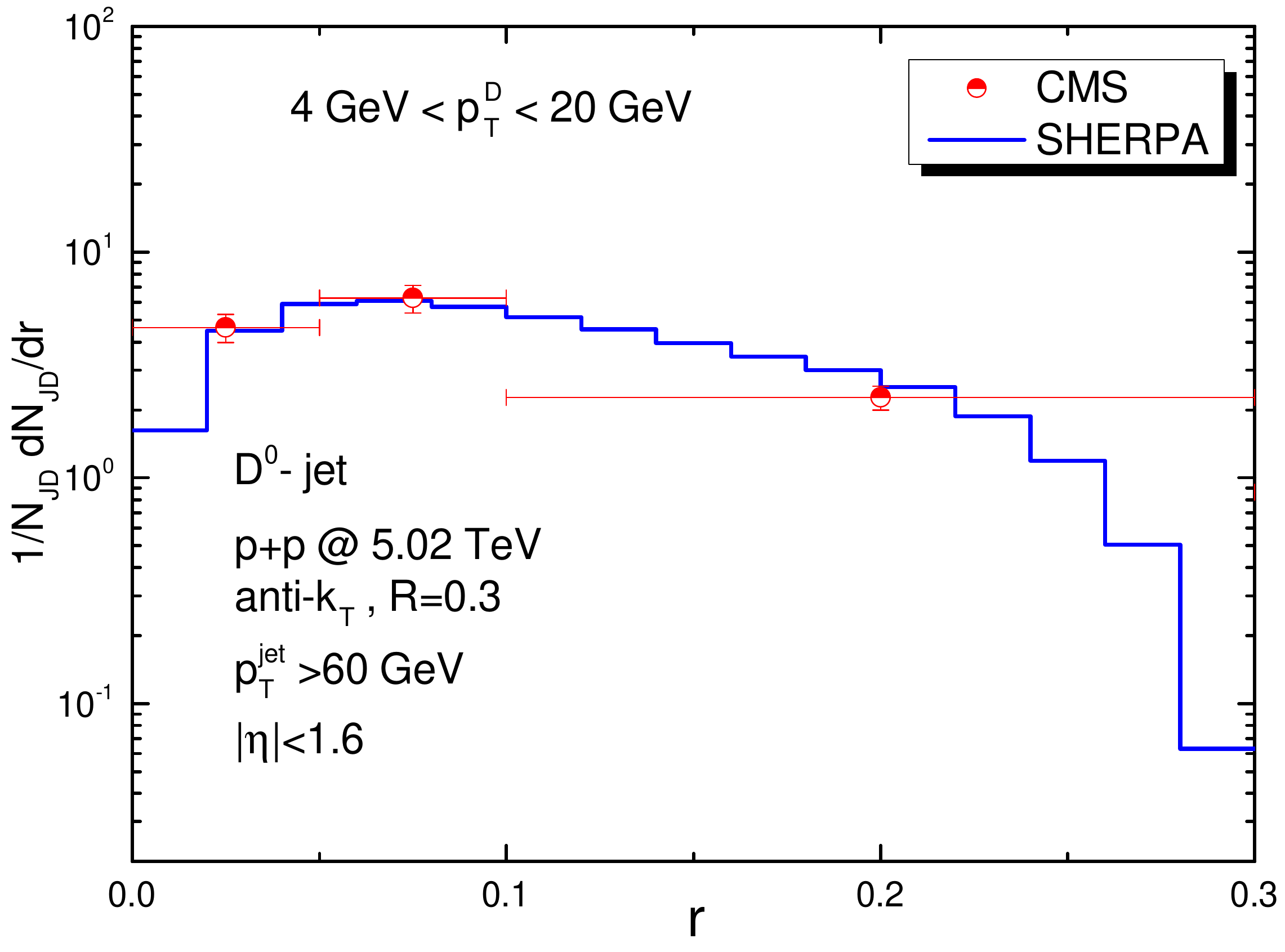, width=3.4in,height=3.in,angle=0, clip=}}
 \subfigure[]{
  \epsfig{file=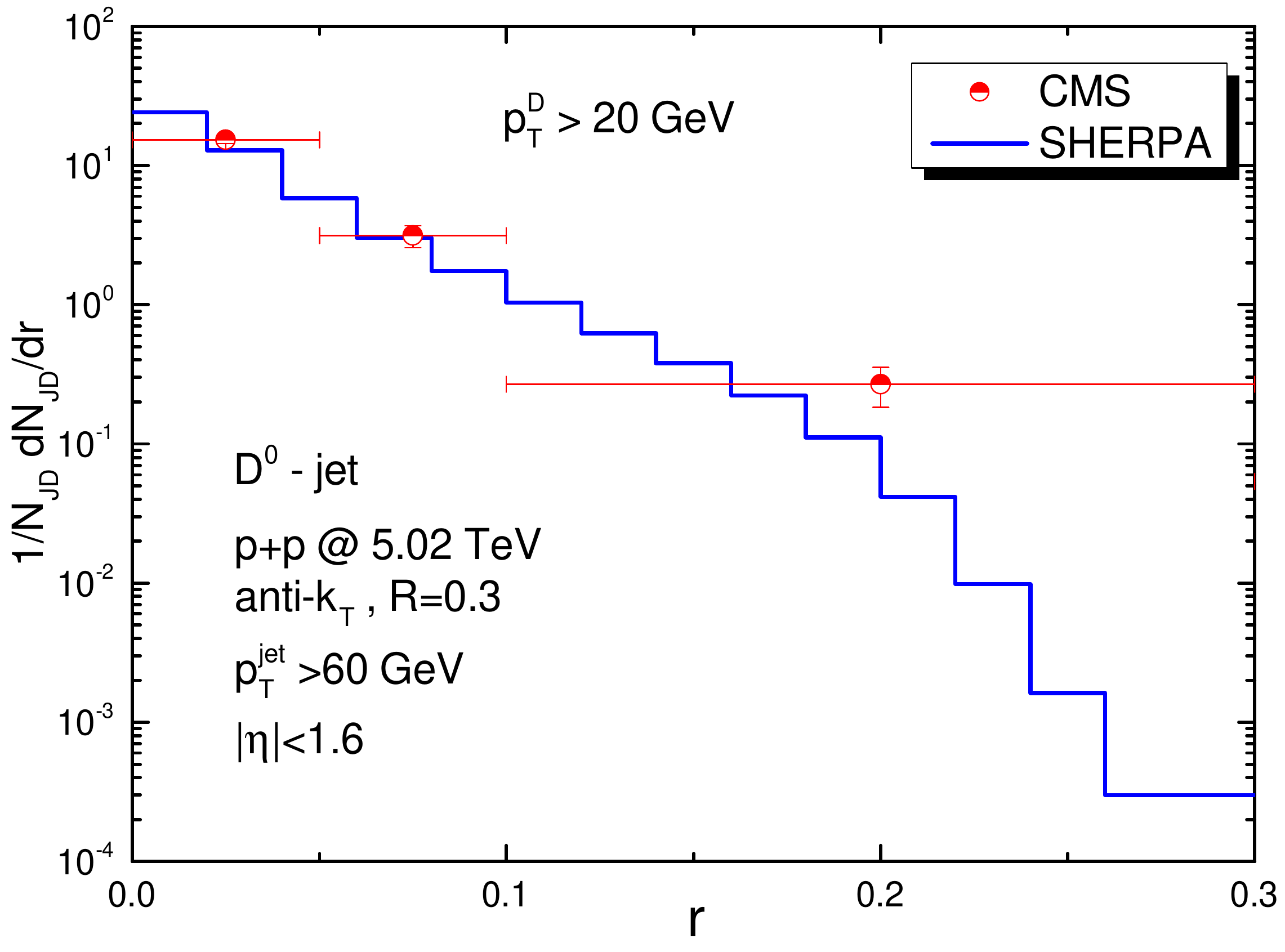,width=3.4in,height=3.in,angle=0, clip=}}
\vspace*{-.1in}
\hspace*{0.2in}
\caption{(Color online) The normalized radial distributions of $D^0$ meson in jets as a function of the angular distance from the jet axis in p+p collisions at $5.02$~TeV provided by SHERPA are compared with CMS data. The jets are selected at $|\eta^{jet}|<1.6$ and with transverse momentum $p^{jet}_{T}>60$~GeV, the proposed $p_T$ constrain for $D^0$ meson are (a) 4~GeV $<p_T<$ 20~GeV and (b) $p_T>$ 20~GeV.}
\label{fig:dndrpp}
\end{center}
\end{figure}

\begin{figure}[!t]
\begin{center}
\vspace*{-0.2in}
\hspace*{-.1in}
\includegraphics[width=3.3in,height=3.1in,angle=0]{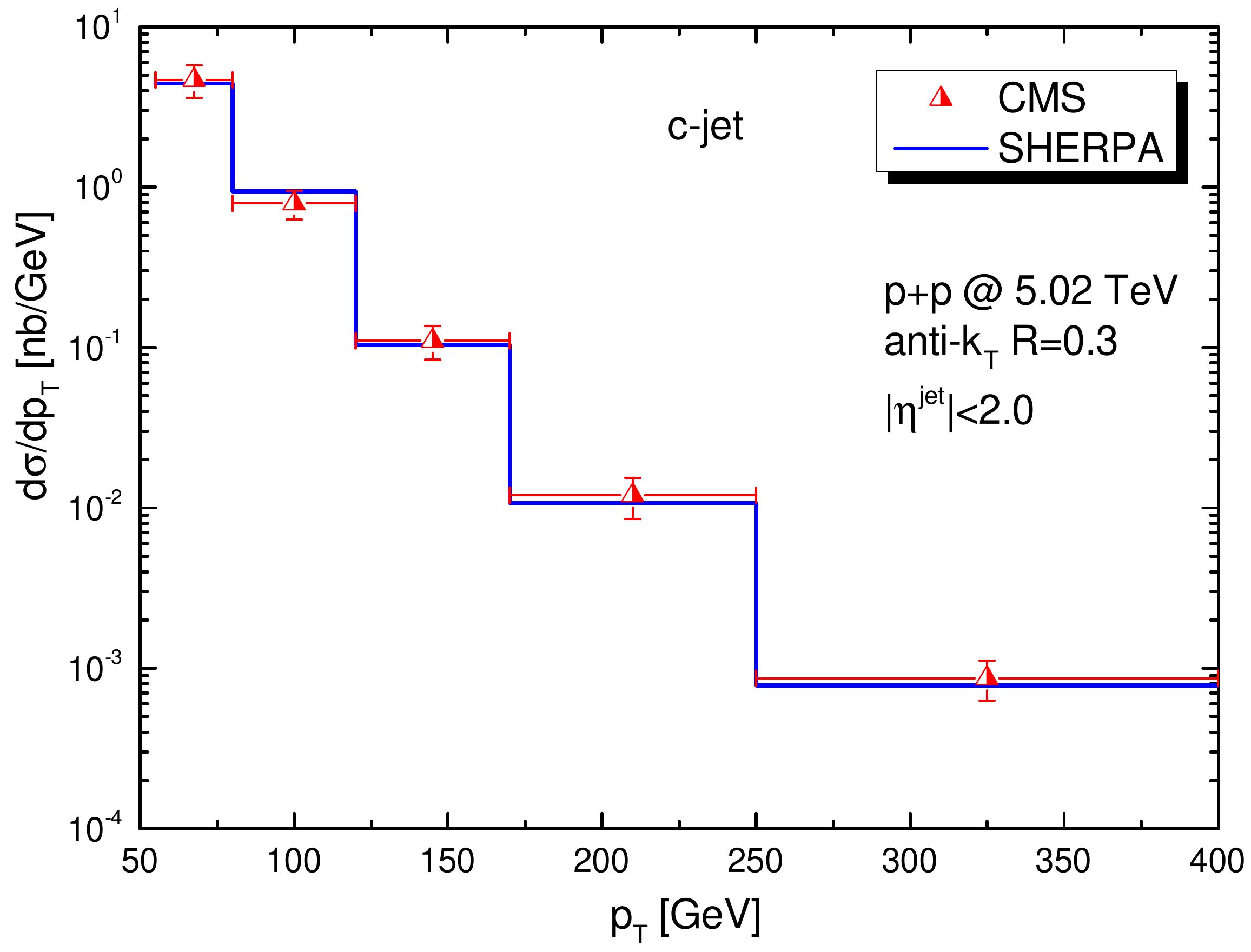}
\vspace*{-.2in}
\caption{(Color online) The c-jet cross section versus c-jet transverse momentum in p+p collisions at $5.02$~TeV with $R=0.3$ and $|\eta^{jet}|<2.0$ and compared with CMS data~\cite{Sirunyan:2016fcs}.}
\label{fig:dsdpt-pp}
\end{center}
\end{figure}

\par The selected jets which reconstructed with radius parameter $\Delta R=\sqrt{(\Delta \phi)^2+(\Delta \eta)^2}=0.3$ must satisfy $p^{jet}_T>60$~GeV and $|\eta^{jet}|<1.6$, where $\phi$ and $\eta$ are the azimuthal angle and the pseudorapidity of the particles. $r=\sqrt{(\Delta\phi_{JD})^2+(\Delta\eta_{JD})^2}$ is defined by the relative azimuthal angle $\Delta\phi_{JD}$ and relative pseudorapidity $\Delta\eta_{JD}$ between the $D^0$ meson and the jet axis. We are able to see that SHERPA could well describe the CMS data~\cite{CMS:2018ovh} measured in p+p collisions both at low $D^0$ meson $p_T$ ($4$~GeV$<p^D_T<20$~GeV) and high $D^0$ meson $p_T$ ($p^D_T>20$~GeV) as shown in Fig.~\ref{fig:dndrpp}. We also note that the shapes of $1/N_{JD}dN_{JD}/dr$ for the two $p^D_T$ range are very different. It is understood that, the jets containing high $p_T$ charm quark (later fragments into $D^0$ meson in the hadronization process) are more likely produced in the initial hard scattering, and the charm quark is still near the jet axis even though after the soft radiation, this is to say, the power-law for the $r$ distribution holds. As for the low $p_T$ $D^0$ meson, charm quarks can be produced by the gluon splitting (GSP) processes during the initial or final state parton shower and unlikely lead the high $p_T$ jets with $p_T^{jet}>60$~GeV. The simulation by SHERPA is also compared with the CMS measurement on c-jet~\cite{Sirunyan:2016fcs} in p+p collisions at $5.02$~TeV shown in Fig.~\ref{fig:dsdpt-pp} indicating that SHERPA could provide a good p+p baseline.

\section{The framework of in-medium parton evolution}
\label{sec:framework}
To implement the parton propagation in the QGP, we treat the p+p events produced by SHERPA with vacuum parton shower as input, and then simulate the subsequent in-medium jet evolution. We initialize the spacial distribution of production vertices via a MC glauber model and assume that all of the partons move like the classical particle. The modified discrete Langevin equations have been considered for the updating of the position and momentum of heavy quarks~\cite{Moore:2004tg,Cao:2013ita,Dai:2018mhw,Wang:2018gxz},
\begin{align*}
\vec{x}(t+\Delta t)&=\vec{x}(t)+\frac{\vec{p}(t)}{E}\Delta t \tag{1}\\
\vec{p}(t+\Delta t)&=\vec{p}(t)-\Gamma(p)\vec{p} \Delta t+\vec{\xi}(t)-\vec{p}_g\tag{2}
\end{align*}
where $\Delta t$ is the time step of the simulation,  and $\Gamma$ the drag coefficient. $\vec{\xi}(t)$ is the stochastic term representing the Gaussian random kicks by the constituents in such thermal medium, which satisfies $\left \langle \xi^i(t)\xi^j(t') \right \rangle =\kappa \delta^{ij}\delta(t-t')$, where $\kappa$ is the diffusion coefficient. Based on the so called Einstein relation, the relationship between $\Gamma$ and $\kappa$ could be expressed as $\kappa=2ET\Gamma=\frac{2T^2}{D_s}$, where $D_s$ denoting the spacial diffusion coefficient is approximatively fixed at $2\pi TD_s=4$ in our simulations according to the Lattice QCD calculation~\cite{Francis:2015daa}. The last term in Eq.(2) represents the modification due to the medium-induced gluon radiation which is sampled based on the higher-twist spectra~\cite{Guo:2000nz,Zhang:2003yn,Zhang:2003wk,Majumder:2009ge}:
\begin{align*}
\frac{dN}{dxdk^{2}_{\perp}dt}=\frac{2\alpha_{s}C_sP(x)\hat{q}}{\pi k^{4}_{\perp}}\sin^2(\frac{t-t_i}{2\tau_f})(\frac{k^2_{\perp}}{k^2_{\perp}+x^2M^2})^4\tag{3}
\end{align*} 
where $x$ and $k_\perp$ are the energy fraction and transverse momentum of the radiated gluon, the last quadruplicate term represents the mass effect of heavy quarks. $C_s$ is the quadratic Casimir in color representation, and $P(x)$ the splitting function in vacuum~\cite{Wang:2009qb}, $\tau_f=2Ex(1-x)/(k^2_\perp+x^2M^2)$ the gluon formation time. $\hat{q} \propto q_0(T/T_0)^3$ is the jet transport parameter~\cite{Chen:2010te}, where $T_0$ is the highest temperature in the most central A+A collisions, and $q_0=1.2$~GeV$^2$/fm is determined by a global extraction of the single hadron production in Pb+Pb collisions at the LHC energy ($q_0=0.5$~GeV$^2$/fm in Au+Au at the RHIC energy)~\cite{Ma:2018swx}. This spectra has also been used to simulate the radiative energy loss of light quarks and gluon. 

Please note the correction term accounted for medium-induced gluon radiation will destroy the fluctuation-dissipation relation. In principle, it will prevent the heavy quark to reach thermal equilibrium in a static medium after evolving for a sufficient long time. However, as an effective way to include the contribution of the radiative energy loss, it is widely used in the simulation with a lower energy cut of the radiated gluon to take into account of the detailed balance between gluon absorption and radiation~\cite{Cao:2013ita,Cao:2018ews}, in this work, we impose the lower energy cut to be $E_0=\mu_D=\sqrt{4\pi\alpha_s}T$ where $\mu_D$ is the Debye screening mass. This treatment makes it possible for the heavy quark to archive thermal equilibrium~~\cite{Cao:2013ita}. To take into account the collisional energy loss of light quarks and gluon, the result calculated under the Hard Thermal Loop (HTL) approximation~\cite{Neufeld:2010xi} $\frac{dE^{coll}}{dt}=\frac{\alpha_{s}C_s\mu_{D}^{2}}{2}ln{\frac{\sqrt{ET}}{\mu_D}}$ has been used to simulate the collisional energy loss in this work. The background hydrodynamic profile of the expanding bulk QGP medium is provided by the smooth iEBE-VISHNU hydro code~\cite{Shen:2014vra}. The parton evolution stops when the local temperature is under $T_c=165$~MeV.

\section{Results and discussions}
\label{sec:results}
\begin{figure}[!t]
\begin{center}
\vspace*{-0.2in}
\hspace*{-.1in}
\subfigure[]{
  \epsfig{file=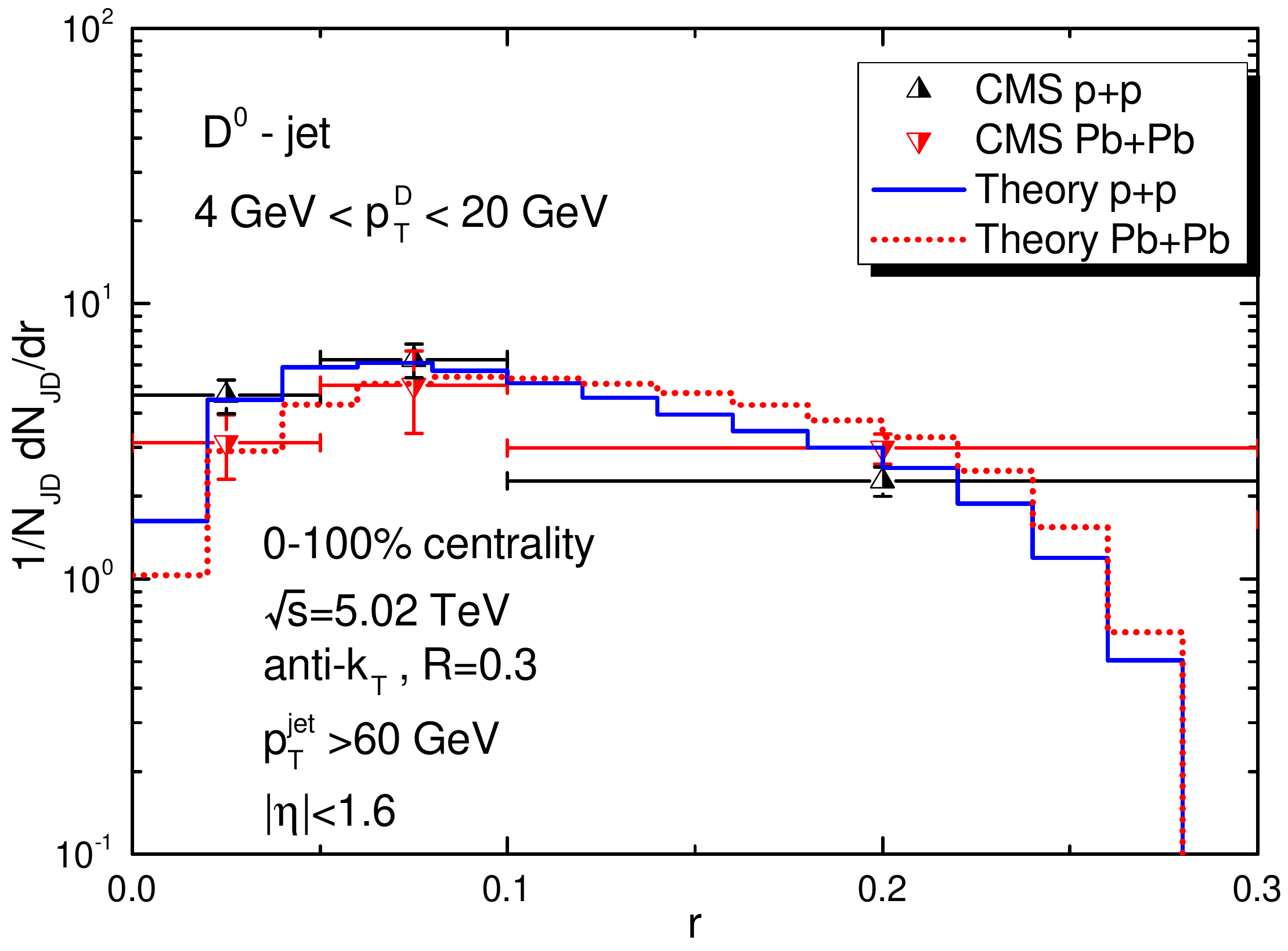, width=3.4in,height=3.in,angle=0, clip=}}
 \subfigure[]{
  \epsfig{file=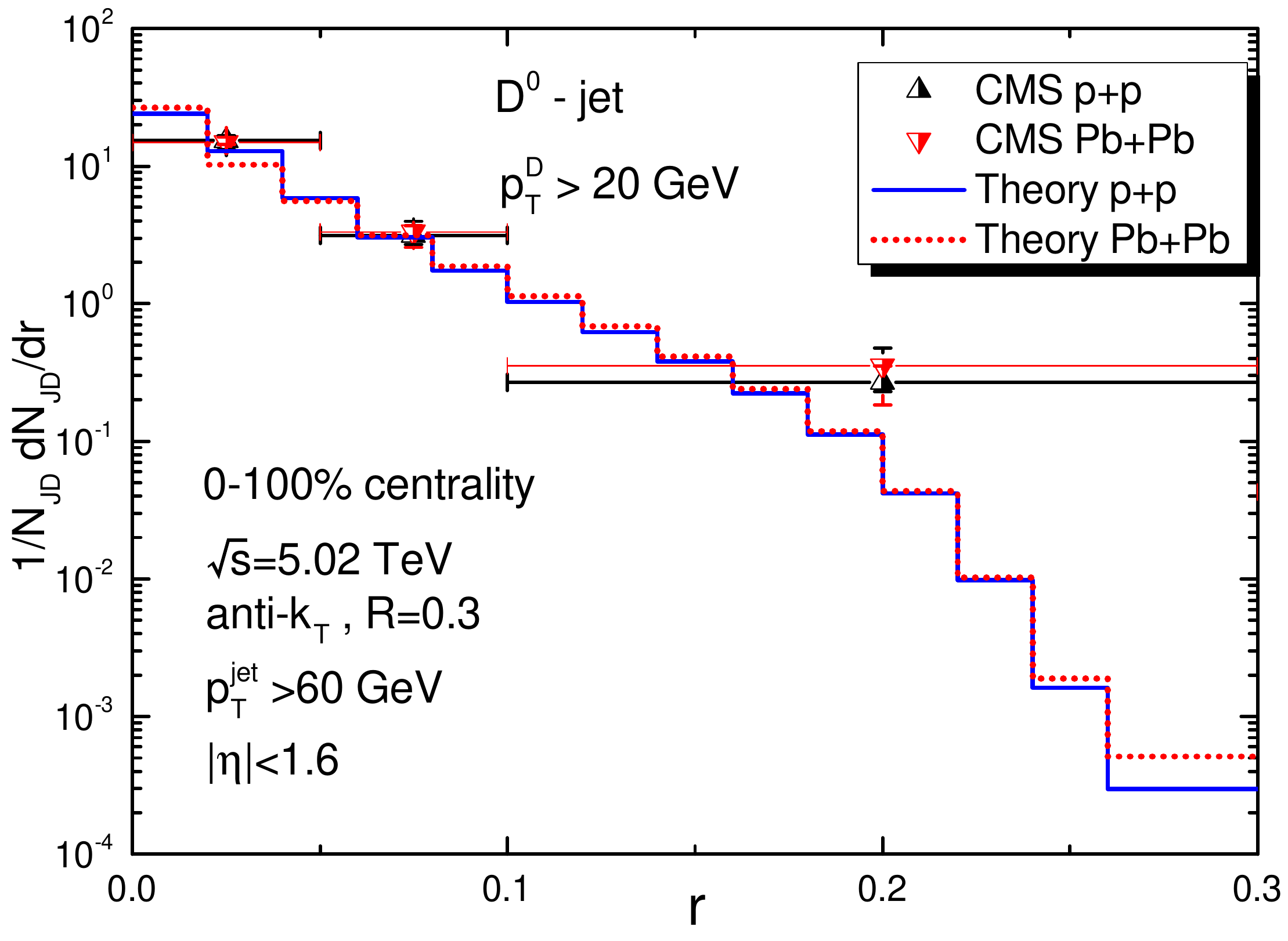,width=3.4in,height=3.in,angle=0, clip=}}
\vspace*{-.1in}
\hspace*{.2in}
\caption{(Color online) Radial distributions of $D^0$ meson in jets at (a) $4$~GeV $< p^D_T < 20$~GeV and (b) $p^D_T > 20$~GeV  both for p+p and Pb+Pb are compared to the CMS data~\cite{CMS:2018ovh}. }
\label{fig:ppAA}
\end{center}
\end{figure}  

\begin{figure}[!t]
\begin{center}
\vspace*{-0.2in}
\hspace*{-.1in}
\subfigure[]{
  \epsfig{file=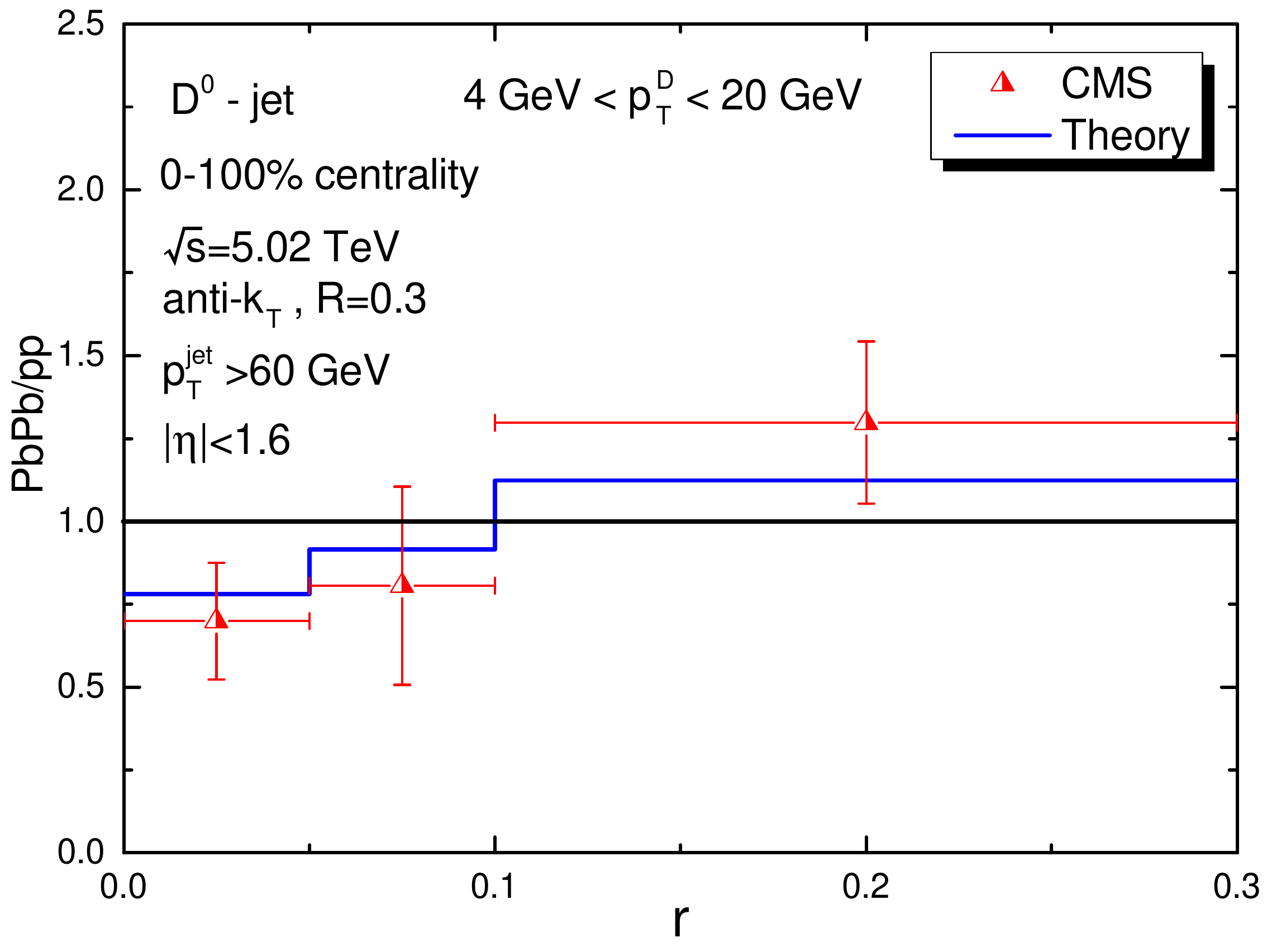, width=3.4in,height=3.in,angle=0, clip=}}
 \subfigure[]{
  \epsfig{file=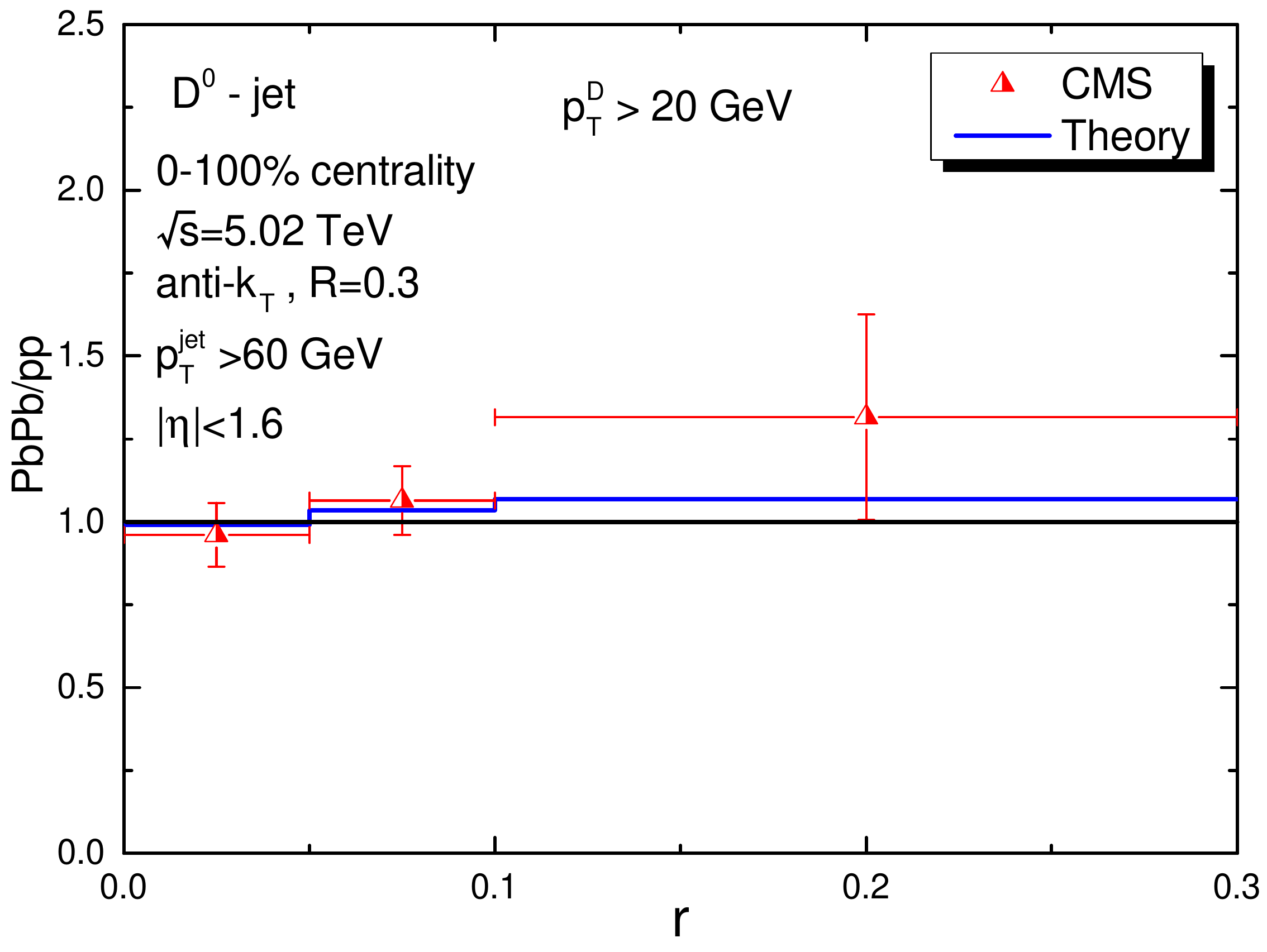,width=3.4in,height=3.in,angle=0, clip=}}
\vspace*{-.1in}
\hspace*{.2in}
\caption{(Color online) Ratios of the radial distribution of $D^0$ meson in jets of Pb+Pb to p+p in the two $p_T$ range: (a) $4$~GeV $< p^D_T < 20$~GeV and (b) $p^D_T > 20$~GeV are compared with the CMS data~\cite{CMS:2018ovh}.}
\label{fig:ratio}
\end{center}
\end{figure}  

\begin{figure}[!t]
\begin{center}
\vspace*{-0.2in}
\hspace*{-.1in}
\includegraphics[width=3.3in,height=3.0in,angle=0]{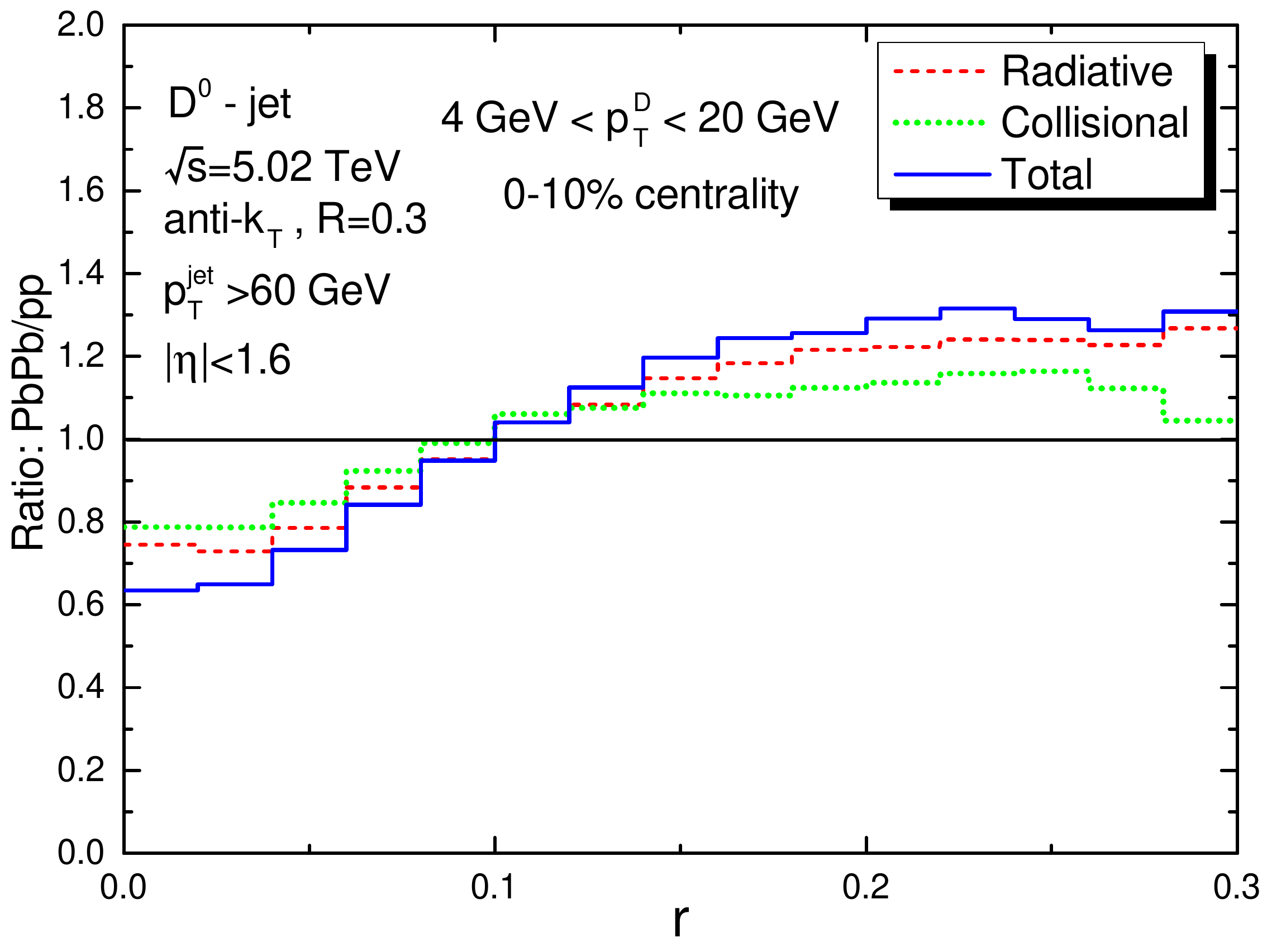}
\vspace*{-.2in}
\caption{(Color online) The ratio of $D^0$ meson radial distribution in jets of $0-10\%$ Pb+Pb to p+p, when only collisional interaction considered(green dotted line), only radiative process considered(red dashed line), and both of them considered(blue line) for charm quarks.}
\label{fig:ratio-radcol}
\end{center}
\end{figure}

\begin{figure}[!t]
\begin{center}
\vspace*{-0.2in}
\hspace*{-.1in}
\subfigure[]{
  \epsfig{file=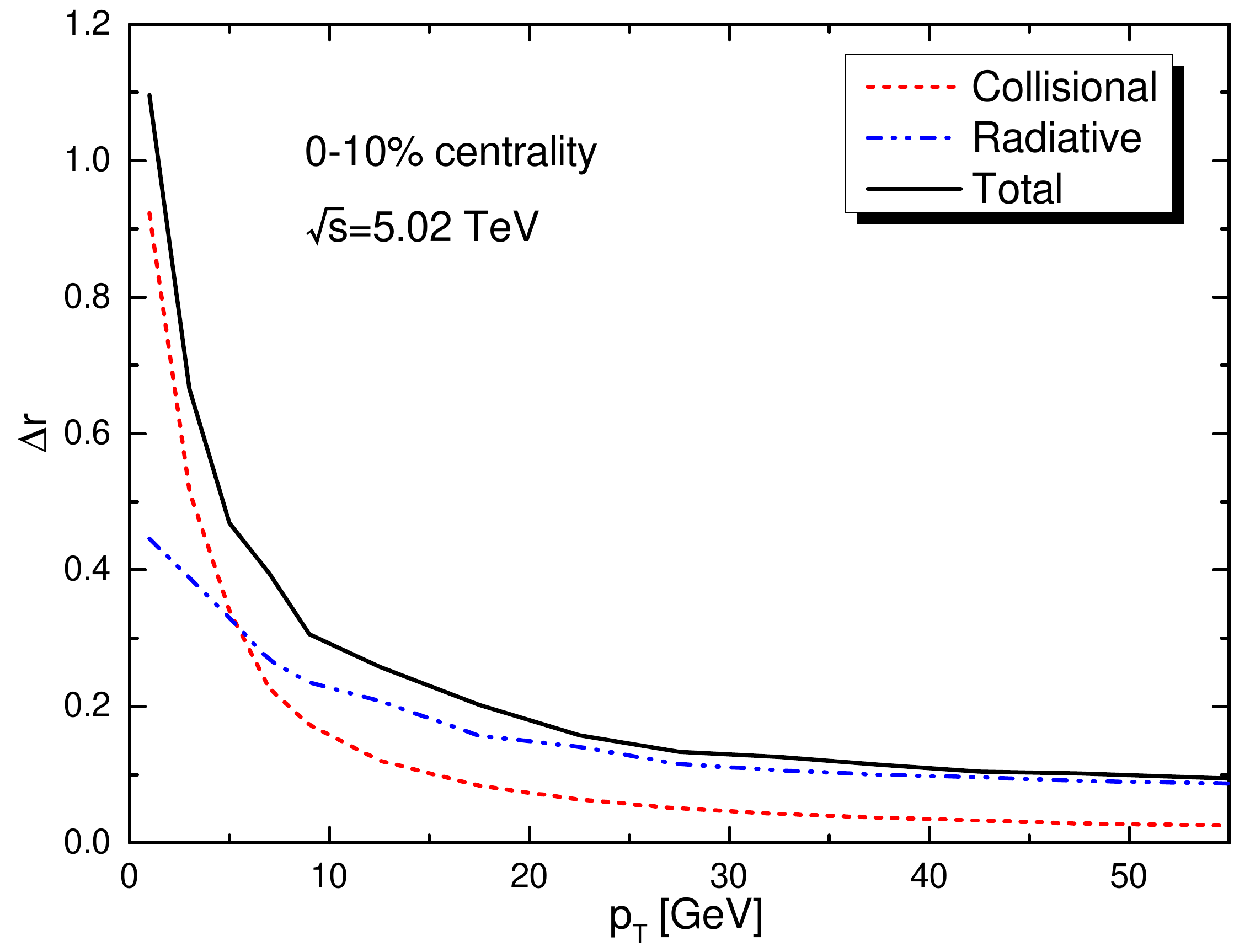, width=3.4in,height=3.in,angle=0, clip=}}
 \subfigure[]{
  \epsfig{file=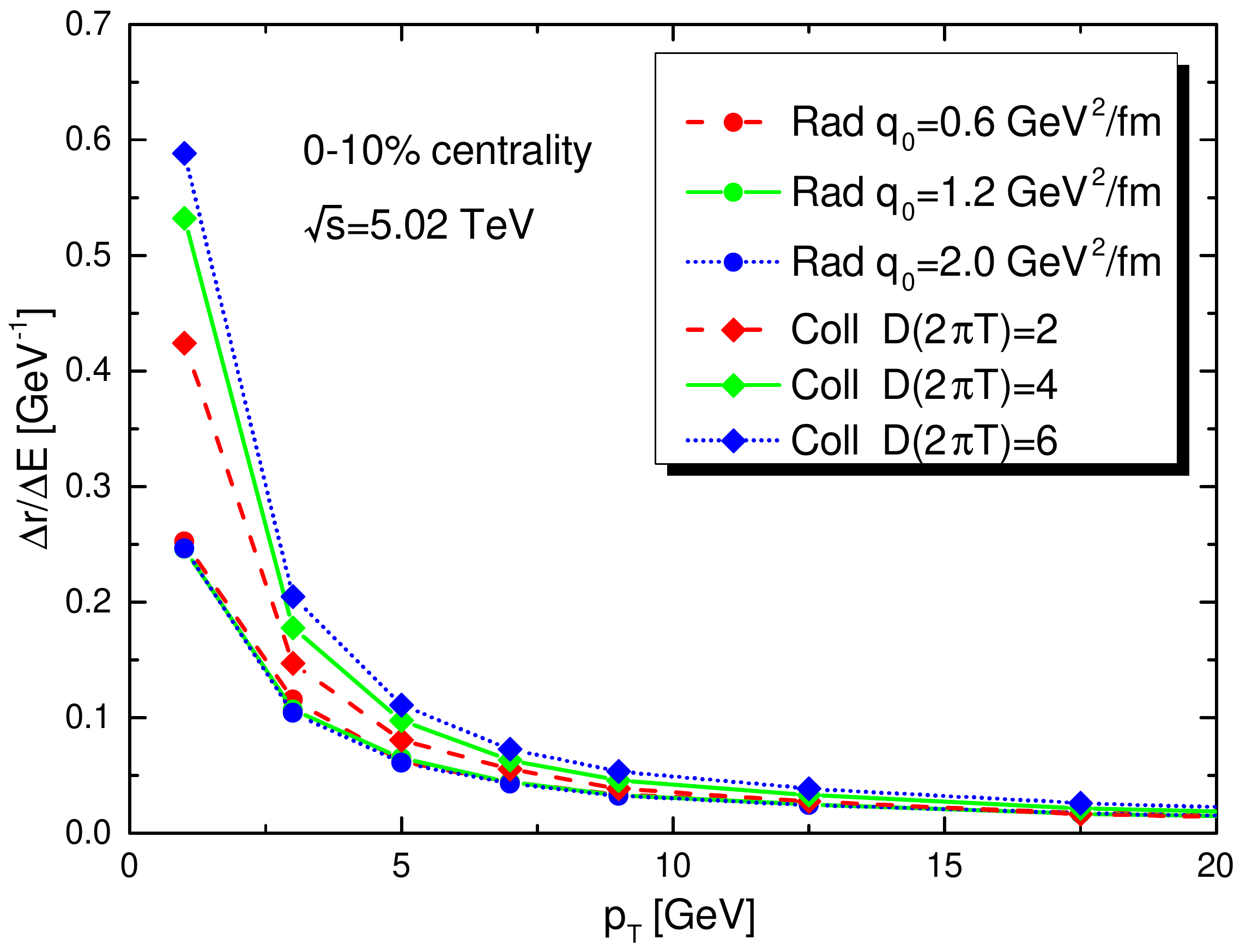,width=3.4in,height=3.in,angle=0, clip=}}
\vspace*{-.1in}
\hspace*{.2in}
\caption{(Color online) (a) Angular deviation of charm quark relative to it's initial direction versus transverse momentum, the net effect from collisional process (red) and radiative process (blue) as well as the total (black) are shown. (b) Comparison of the ratio of angular deviation to energy loss as a function of transverse momentum, both for collisional and radiative, are shown with different parameters.}
\label{fig:drdt}
\end{center}
\end{figure}  

\begin{figure}[!t]
\begin{center}
\vspace*{-0.1in}
\hspace*{-0.1in}
\includegraphics[width=3.4in,height=3.in,angle=0]{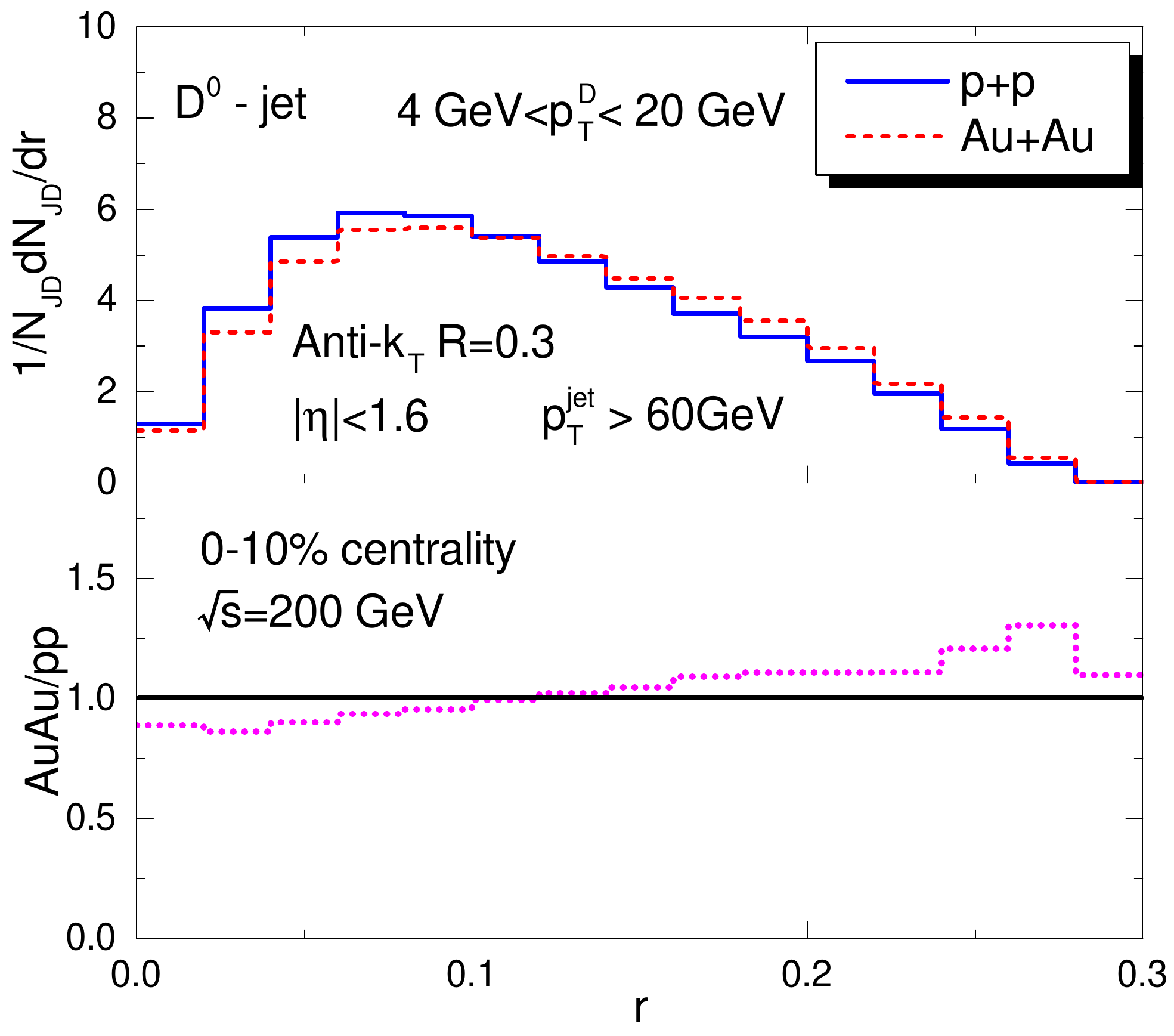}
\vspace*{-.2in}
\caption{(Color online) The normalized radial distribution of $D^0$ meson in jets in p+p and $0-10\%$ Au+Au collisions at $\sqrt{s_{NN}}=200$~GeV as well as the ratio of Au+Au to p+p.}
\label{fig:ratio-rhic}
\end{center}
\end{figure}

\begin{figure}[!t]
\begin{center}
\vspace*{-0.1in}
\hspace*{-.1in}
\subfigure[]{
  \epsfig{file=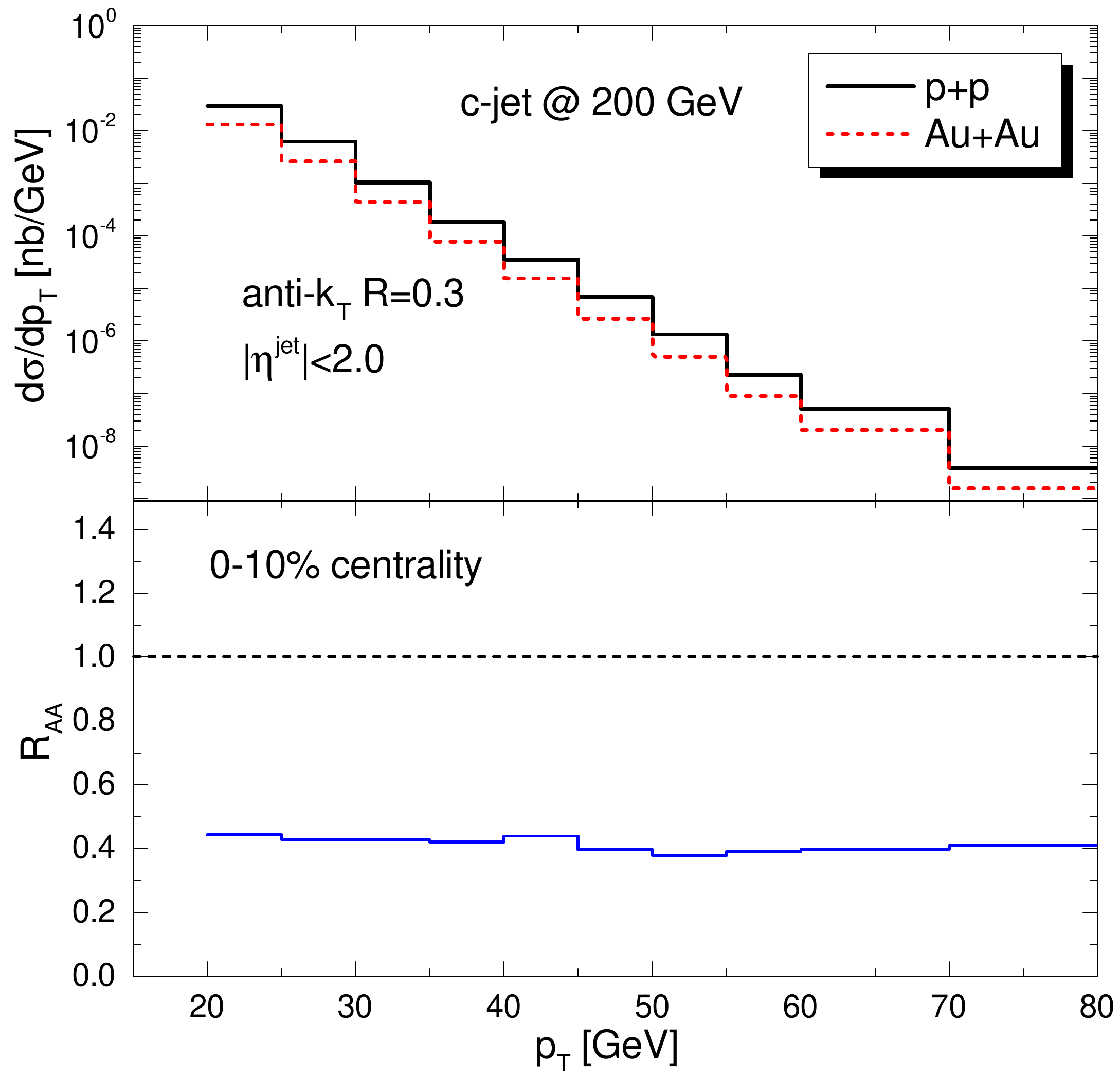, width=3.4in,height=3.in,angle=0, clip=}}
 \subfigure[]{
  \epsfig{file=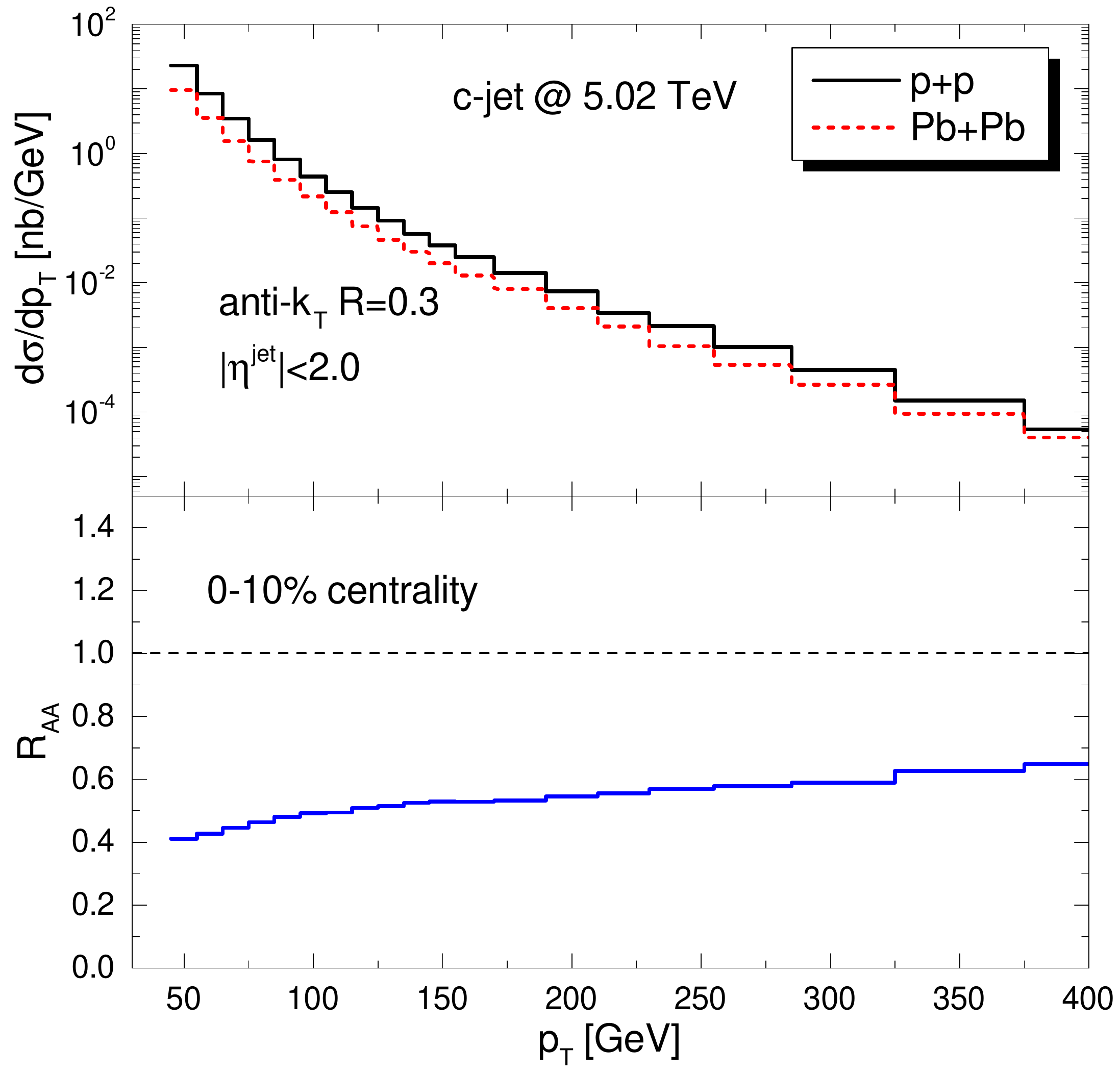,width=3.4in,height=3.1in,angle=0, clip=}}
\vspace*{-0.1in}
\hspace*{-.1in}
\caption{(Color online) (a) The c-jet cross section versus transverse momentum in (a) p+p and $0-10\%$ Au+Au collisions at 200~GeV, (b) p+p and $0-10\%$ Pb+Pb collisions at 5.02~TeV, as well as the nuclear modification factor $R_{AA}(p_T)$. The charm jets are reconstructed with cone size $R=0.3$ and $|\eta^{jet}|<2.0$. }
\label{fig:dsdpt-AA}
\end{center}
\end{figure}

 In Fig~\ref{fig:ppAA}, we show our calculated results of the $D^0$ meson radial distribution in jets in the $0-100\%$ Pb+Pb collisions at $\sqrt{s_{NN}}=5.02$~TeV and find that our results can provide quite decent descriptions on the CMS data~\cite{CMS:2018ovh}. At low $D^0$ meson $p_T$ ($4-20$~GeV), a visible shift towards large $r$ is found in the normalized distribution $\frac{1}{N_{JD}}\frac{dN_{JD}}{dr}$ in Pb+Pb collisions relative to the p+p baseline.  Whereas, for high $p_T$ $D^0$ mesons, no significant modification is observed for the $D^0$-jet correlation in the Pb+Pb collisions. It indicates that the low $p_T$ $D^0$ mesons are obviously farther away from the jet axis due to the in-medium interaction in Pb+Pb collisions. To make a quantitative description of this effect, the ratio of the radial distribution of $D^0$ meson in jets of Pb+Pb to p+p has also been estimated in Fig.~\ref{fig:ratio}. The ratios increase with $r$ and the magnitude are well consistent with the CMS measurement both for low $p_T$ and high $p_T$ $D^0$ mesons.

\par The diffusion of charm quarks in the hot and dense QGP medium results from the joint effect of elastic and inelastic processes. In our framework, the Langevin mechanism (Brownian motion) and induced gluon radiation describe the elastic and inelastic interaction between the charm quarks and the constituents of the QGP medium respectively. On the one hand, the stochastic term $\vec{\xi}$ in Eq.(2) representing the elastic collisions is the main source of the charm quark diffusion at not very large $p_T$ within Langevin mechanism. A mass of random kicks from medium cause the spacial diffusion and also change the moving direction of charm quarks, and subsequently the charm quarks near the jet axis will diffuse to larger $r$. On the other hand, the last term $-\vec{p}_g$ in Eq.(2) denotes the recoil momentum of charm quarks caused by the medium-induced gluon radiation. Due to its significant contribution to the charm quark energy loss, this radiative process would also play an important role in the charm diffusion in the QGP medium.
\par To estimate the net effect from collisional process and radiative process, we show the radial distribution modification of $D^0$ meson in jets in Pb+Pb collisions result from different mechanism in Fig.~\ref{fig:ratio-radcol}. We find that both of collisional and radiative mechanisms play important role in the diffusion of charm quark inside the jets, but the effect from latter is more pronounced. We also note that the collisional process shifts this distribution to moderate $r$ ($0.1-0.25$), but has little effect at the region $r\sim 0.3$, shows a quite different behavior as compared to the radiative process which strongly enhances the distribution even to the boundary of the jet. We can make sense as following: elastic collisions diffuse the charm quarks closing to the jet axis to larger radius inside the jet cone, but the stronger effect from inelastic collisions diffuse them even to outside of the jets.

\par To further study the diffusion effects result from the collisional and radiative interaction in the QGP, we define $\Delta r=\sqrt{(\phi^c-\phi^c_0)^2+(\eta^c-\eta^c_0)^2}$ to quantify the angular deviation of charm quark from its original moving direction during the propagation in medium, where $\phi^c_0$ and $\eta^c_0$ are the initial azimuthal angle and initial pseudorapidity of charm quarks. We show in Fig.~\ref{fig:drdt} (a) the $\Delta r$ as a function of the initial transverse momentum of charm quark. At very low $p_T$ (0-5~GeV), we observe a large $\Delta r$ which is dominated by collisional process. As $p_T$ goes higher, the diffusion effect due to elastic collision falls quickly and then below that due to the radiation when $p_T >  5$~GeV and also the total effects fall with increasing $p_T$. The $p_T$ dependence of the charm angular deviation explains the different modification between low and high $p^D_T$ range in the CMS measurement~\cite{CMS:2018ovh}, and it also indicates that the low $p_T$ $D^0$ meson in jets may act as a new sensitive probe to the disentangling of collisional and radiative mechanisms when the future precise measurements are available. Even though the angular deviation caused by radiation is also small at high $D^0$ meson $p_T$, a none-zero $\Delta r$ due to radiation still exists and dominates the total diffusion effect even at $p_T\sim 50$~GeV. Beyond that, to estimate the diffusion strength per unit energy loss in the two mechanisms, the ratios of angular deviation to the energy loss as a function of D meson $p_T$ are shown in Fig.~\ref{fig:drdt} (b). Distinctly, the ratios $\Delta r/\Delta E$ decline rapidly due to the increasing energy loss when $p_T$ increases. However, The $\Delta r/\Delta E$ of collisional interaction is much larger than that of radiative mechanism especially at $p_T < 5$~GeV. In addition to this, we discover that the $\Delta r/\Delta E$ of radiative mechanism is not sensitive to the variance of $q_0$, which controls the strength of in-medium gluon radiation,  because that $\Delta r$ and $\Delta E$ are all proportional to $\hat{q}$ as implied in the Eq.(3), which is in contrast with the collisional mechanism. The differences in the angular deviation between these two mechanisms may be helpful to disentangle the collisional and radiative contribution for heavy quarks in-medium interaction.

\par We also present the prediction on the radial distribution of the $D^0$ meson in jets in p+p and central $0-10\%$ Au+Au collisions at the RHIC energy shown in Fig~\ref{fig:ratio-rhic}. Using the same configuration in the jet reconstruction as above, we can also observe a visible modification of $D^0$ radial distribution in jets in $0-10\%$ Au+Au collisions at 200~GeV, but it seems weaker than that in the $0-10\%$ Pb+Pb collisions at 5.02~TeV shown in Fig~\ref{fig:ratio-radcol}. This mainly due to the fact that both $\kappa$ and $\hat{q}$ which control the collisional and radiative strength are all proportional to the $T^3$, hence the diffusion effects of charm quarks strongly dependent on the local temperature of the formed medium during their propagation. 
\par For completeness, we also give predictions on the nuclear modification of c-jet production. Shown in Fig~\ref{fig:dsdpt-AA}, we calculate the differential cross section of c-jet as a function of $p_T$ in p+p and A+A collisions both at the RHIC and the LHC energy respectively, as well as the predicted nuclear modification factor $R_{AA}$ of c-jet ($R_{AA}$ of D-jet is very close to that of c-jet). In $0-10\%$  Au+Au collisions at $200$~GeV, the $R_{AA}\sim 0.4$ shows a weak dependence on c-jet $p_T$. At the Pb+Pb collisions at $5.02$~TeV, the $R_{AA}$ varies from 0.4 to 0.6 with increasing $p_T$ at $50-400$~GeV.

\section{Summary}
\label{sec:summ}
 In summary, we study the charm diffusion effects, which differ from the perspective of heavy quarks energy loss, on the $D^0$-jet angular correlations within a Monte Carlo simulation method. Our simulated results can provide a quite decent description on the experimental data within the uncertainty both in p+p and Pb+Pb collisions. We estimate the net effect on the charm diffusion from collisional and radiative mechanism and demonstrate the $p_T$ dependence of this diffusion effect. We find that collisional process has significant effects at low $p_T$, especially dominates at $0-5$~GeV, and the radiative process has a non-zero effect even at high $p_T\sim 50$~GeV. The total diffusion effect decreases with $D^0$ meson $p_T$ which explains the significant modification at low $D^0$ meson $p_T$ measured in experiment. Besides, for unit energy loss, the collisional interaction shows much stronger angular deviation than radiative interaction at $p_T<5$~GeV. The strong diffusion effects of low $p_T$ charm quarks relative to jets may act as a sensitive probe to the distinction between collisional and radiative interaction, and the prospective measurements would provide more precise estimations on the heavy quarks diffusion coefficient and also give additional constrains for the current theoretical models on heavy quark energy loss. As our theoretical predictions, we present the $D^0$ meson radial distribution in jets in p+p and $0-10\%$ Au+Au collisions at $\sqrt{s_{NN}}=200$~GeV at the RHIC, a visible modification is observed in our results. We also estimate the nuclear modification factor for charm jet in $0-10\%$ Au+Au collisions at $\sqrt{s_{NN}}=200$~GeV at the RHIC and $0-10\%$ Pb+Pb collisions at $\sqrt{s_{NN}}=5.02$~TeV at the LHC as predictions for the future measurements.

\section*{Acknowledgments}
\label{sec:ack}
We would like to thank Marta Verweij for helpful discussion about the CMS measurement. This research is supported by the NSFC of China with Project Nos. 11435004, 11805167, and partly supported by China University of Geosciences (Wuhan) (No. 162301182691).

\end{document}